\documentclass[pdflatex,sn-mathphys-num]{sn-jnl}% Math and Physical Sciences Numbered Reference Style

\usepackage{graphicx}%
\usepackage{multirow}%
\usepackage{amsmath,amssymb,amsfonts}%
\usepackage{amsthm}%
\usepackage{mathrsfs}%
\usepackage[title]{appendix}%
\usepackage{xcolor}%
\usepackage{textcomp}%
\usepackage{manyfoot}%
\usepackage{booktabs}%
\usepackage{algorithm}%
\usepackage{algorithmicx}%
\usepackage{algpseudocode}%
\usepackage{listings}%
\usepackage{tikz}
\usetikzlibrary{arrows.meta, positioning, calc, fit}

\usepackage{listings}
\usepackage{xcolor}

\usetikzlibrary{positioning,fit,calc,arrows.meta,backgrounds,shapes.geometric}
  \usepackage{amsmath,amssymb}
  \usepackage{subcaption}   % only for the figure* wrapper below
\lstdefinelanguage{json}{
    basicstyle=\ttfamily\normalsize ,
    numbers=none,
    breaklines=true,
    breakatwhitespace=false,
    showstringspaces=false,
    frame=single,
    rulecolor=\color{gray!50},
    backgroundcolor=\color{gray!5},
    string=[s]{"}{"},
    stringstyle=\color{blue!60!black},
    commentstyle=\color{gray},
    morestring=[b]',
}

\theoremstyle{thmstyleone}%
\theoremstyle{thmstyletwo}%

\theoremstyle{thmstylethree}%

\begin{document}

\title[Benchmarking CLEF and CLEF-Flash for Breast Ultrasound Classification]{Can Multimodal Decision Models Classify Breast Ultrasound Images? Benchmarking CLEF and CLEF-Flash Across Prompts and Datasets}

\author*[1]{\fnm{Tran Cao} \sur{Minh}}\email{trancaominhkg@gmail.com}

\author[2]{\fnm{Hai} \sur{Tran}}
\email{haits@hcmue.edu.vn}

\author[3,4]{\fnm{Ha Minh} \sur{Tan}}
\email{tanhm@uit.edu.vn}

\affil[1]{%
\orgdiv{Research and Development Center},
\orgname{VNSO Technology JSC},
\city{Ho Chi Minh City},
\country{Vietnam}
}

\affil[2]{%
\orgdiv{Department of Information Technology},
\orgname{Ho Chi Minh University of Education},
\city{Ho Chi Minh City},
\country{Vietnam}
}

\affil[3]{%
\orgdiv{Faculty of Information Science and Engineering},
\orgname{University of Information Technology},
\orgaddress{
\city{Ho Chi Minh City},
\country{Vietnam}}
}

\affil[4]{%
\orgname{Vietnam National University},
\orgaddress{
\city{Ho Chi Minh City},
\country{Vietnam}}
}


\abstract{Multimodal decision models return probabilities over predefined alternatives, offering a structured interface for image classification. This study benchmarks inference with CLEF and CLEF-Flash on 2818 breast ultrasound images from Dataset B, BUSI, and BUS-BRA. Dataset B and BUS-BRA use binary benign-malignant classification, whereas BUSI includes normal, benign, and malignant categories. Model parameters remain unchanged, and original images are submitted without user-defined preprocessing. Three prompts provide minimal labels, an explicit classification task, or sonographic criteria. We report accuracy, macro F1, balanced accuracy, macro precision, throughput, and latency. CLEF-Flash with the clinical prompt achieves the highest accuracy on all three datasets: 74.85\% on Dataset B, 66.28\% on BUSI, and 51.63\% on BUS-BRA. On BUSI, however, CLEF with the task-guided prompt achieves the highest macro F1 (64.51\%) and balanced accuracy (68.72\%). CLEF-Flash has lower latency in every matched comparison. Clinical instructions improve CLEF-Flash accuracy relative to the minimal prompt but reduce CLEF accuracy on all datasets. Confusion matrices reveal frequent benign-to-malignant errors, particularly on BUS-BRA, where all configurations fall below the 67.63\% accuracy of a majority-class predictor. These findings demonstrate dataset-and prompt-dependent performance and motivate class-specific evaluation before applying multimodal decision models to breast ultrasound analysis.}

%%===================




\maketitle
\section{Introduction}

Breast ultrasound is widely used for the examination and characterization of breast lesions because it is noninvasive, accessible, and does not involve ionizing radiation. In parallel with its clinical use, artificial intelligence has been extensively investigated for computer aided breast ultrasound analysis, particularly for distinguishing benign from malignant lesions. Deep learning methods based on convolutional neural networks and, more recently, transformer architectures have achieved promising diagnostic performance \cite{dan2024}. However, conventional approaches generally depend on task specific training data, carefully designed training pipelines, and model optimization for a particular dataset or clinical task. These requirements can limit the portability of a trained model across institutions, acquisition devices, and datasets \cite{minh2026swindeeplabv3}. Recent research therefore increasingly considers more generalizable models that can transfer knowledge across tasks and domains.

Foundation models offer an alternative to developing a separate representation for each imaging task. MedCLIP supports prediction through representations learned from medical images and text \cite{wang2022medclip}, while BiomedCLIP extends biomedical vision-language pretraining to millions of scientific image-text pairs \cite{zhang2024biomedclip}. LLaVA-Med explores visual instruction following for biomedical images \cite{li2023llavamed}. These approaches differ in their training objectives and interfaces, but collectively motivate testing how pretrained multimodal systems respond to medical images and task descriptions at inference time.

A related development is the emergence of promptable vision models. The Segment Anything Model showed that a sufficiently large foundation model could perform segmentation on previously unseen image distributions through prompts rather than conventional task specific retraining \cite{kirillov2023sam}. Its adaptation to medical imaging, MedSAM, was subsequently developed using more than one million medical image mask pairs and demonstrated strong generalization across multiple imaging modalities and anatomical targets \cite{ma2024medsam}. These developments have encouraged a broader transition from designing a separate model for every individual task toward interacting with large pretrained models through prompts. At the same time, specialized foundation models are also being developed for breast ultrasound. For example, BUSGen was pretrained on more than 3.5 million breast ultrasound images and demonstrated strong performance after downstream adaptation \cite{yu2026busgen}. Such approaches illustrate the potential of large scale foundation models, but they also continue to rely on extensive medical domain data and additional task adaptation.

More recently, \textit{decision models} have introduced a structured interface for assigning probabilities to predefined alternatives. Cloudflare's CLEF and CLEF-Flash extend this approach to multimodal inputs \cite{cloudflare2026clef}, motivating an evaluation of their inference capabilities for medical image classification.

Despite rapid progress in multimodal foundation models, the capability of these newly introduced decision models for medical image analysis remains largely unexplored. In particular, their performance when applied directly to breast ultrasound images with task instructions is unclear. Breast ultrasound provides a useful setting for investigating this question because lesion appearance, image quality, acquisition equipment, and operator dependent imaging conditions vary across data sources.

To address this gap, we benchmark inference with CLEF and CLEF-Flash across multiple datasets and prompting strategies. The study focuses on the diagnostic predictions obtained from the released models under a common evaluation protocol.

The main contributions of this study are summarized as follows:

\begin{itemize}
    \item We provide an empirical benchmark of inference with CLEF and CLEF-Flash for breast ultrasound classification.

    \item We investigate three prompting strategies with different levels of task and clinical information to characterize the influence of inference instructions on multimodal decision model performance.

    \item We assess performance across three public breast ultrasound datasets covering binary and three class diagnostic settings.
\end{itemize}

The remainder of this paper is organized as follows. Section~\ref{sec:decision_models} introduces decision models and their architecture. Section~\ref{sec:methods} describes the datasets and inference protocol. Sections~\ref{sec:results} and~\ref{sec:discussion} present and discuss the findings, and Section~\ref{sec:conclusion} concludes the paper.

\section{Decision Models}
\label{sec:decision_models}

Decision models are designed to produce structured outputs over a bounded set of alternatives. TypeSafe describes this interface through its Jev System One model \cite{typesafe2026jev}. The formulation is relevant to classification because it exposes an explicit distribution over the requested labels rather than requiring a label to be extracted from free-form text.

Formally, let $x$ denote an input state and let $Q=\{q_1,q_2,\ldots,q_m\}$ denote a set of decision questions. Each question $q_j$ is associated with a predefined answer space $A_j=\{a_{j1},a_{j2},\ldots,a_{jk}\}$. For a \texttt{choice} question, the official TypeSafe specification defines the output as a probability distribution over these alternatives, with the highest-probability option selected as the answer \cite{typesafe2026choice}. Given the state and decision schema, the model estimates

\begin{equation}
\label{eq:decision_probability}
P(a_{ji}\mid x,q_j), \qquad a_{ji}\in A_j,
\end{equation}

where

\begin{equation}
\sum_{a_{ji}\in A_j} P(a_{ji}\mid x,q_j)=1.
\end{equation}

The predicted decision is then obtained as

\begin{equation}
\hat{a}_j=\arg\max_{a_{ji}\in A_j}P(a_{ji}\mid x,q_j).
\end{equation}

The predicted probability of the selected option is $p_{j,\max}=P(\hat{a}_j\mid x,q_j)$, whereas the meaning of the returned \texttt{confidence} field depends on the implementation. For a \texttt{choice} question with $n_j=|A_j|>1$ alternatives, TypeSafe defines confidence as $(p_{j,\max}-1/n_j)/(1-1/n_j)$: it equals zero for a uniform distribution and one when a single option receives all probability mass \cite{typesafe2026confidence}. Cloudflare's released CLEF implementation instead sets \texttt{confidence} to $p_{j,\max}$ rounded to four decimal places in its \texttt{systemone\_answer} function \cite{cloudflare2026clefconfidence}. Thus, the TypeSafe normalization and the CLEF response field have distinct definitions despite their shared name. TypeSafe further specifies that confidence thresholds should be evaluated on data from the intended use case \cite{typesafe2026confidence}.

This formulation differs from traditional classification heads because the decision alternatives are specified at inference time through a typed schema rather than being permanently fixed during supervised training. As a result, the same pretrained model can be applied to multiple decision tasks without retraining a separate classifier for each label set. The output is also directly interpretable as a probability distribution over bounded alternatives, which makes the framework naturally compatible with medical image classification.

\begin{figure*}[t]
    \centering
    \includegraphics[width=\textwidth]{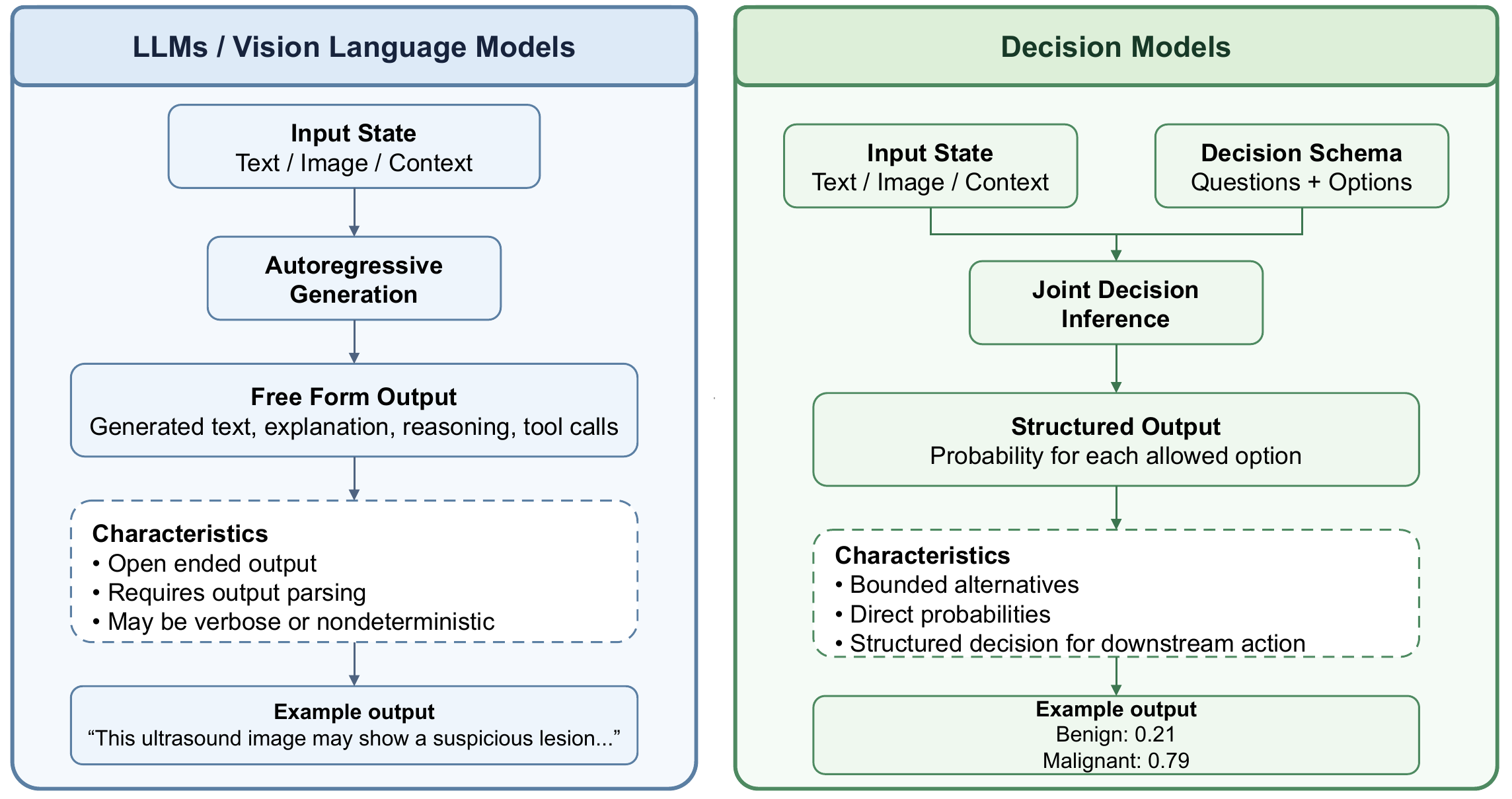}
    \caption{Conceptual comparison between conventional large language or vision language models and decision models. Generative models produce open ended outputs, whereas decision models operate over a predefined decision schema and return structured probabilities for bounded alternatives.}
    \label{fig:llm_vs_decision_models}
\end{figure*}

Figure~\ref{fig:llm_vs_decision_models} illustrates the contrast with autoregressive generation, whose textual outputs may require additional parsing or post processing before use in a classification workflow.

\textbf{Architecture of CLEF and CLEF-Flash}

Figure~\ref{fig:clef_architectures} illustrates the shared architecture of CLEF and CLEF-Flash, including their multimodal backbones and joint schema heads. CLEF and CLEF-Flash share the same backbone-plus-head design, in which a pretrained multimodal language model serves as a frozen encoder and a lightweight decision head replaces the autoregressive language-modeling output \cite{cloudflare2026clef,cloudflare2026clefhf}. CLEF is built on Qwen3.8-27B, whereas CLEF-Flash is built on Qwen3.5-9B. In both models the native vision encoder of the backbone is retained, so images and video frames are converted into visual tokens and processed together with textual tokens in a single sequence. The model input is a serialized record that contains the state $x$ (free text, JSON, images, or video) and the decision schema $Q$. Each question $q_j$ in the schema carries a type (binary \texttt{noul}, nominal \texttt{choice}, or ordinal \texttt{score}), optional natural-language instructions, and a textual description of every admissible option in $A_j$.

Inference proceeds in two stages. In the first stage, the backbone performs a single prefill-only forward pass over the whole record and produces final-layer hidden states $\mathbf{H}\in\mathbb{R}^{T\times d}$, where $T$ is the sequence length and $d$ is the hidden dimension. No token is generated autoregressively. In the second stage, a small transformer module, called the joint schema head, operates on $\mathbf{H}$ through a two-stage attention routing mechanism. First, the representation of each admissible option attends to the state tokens to gather option-specific evidence. Then the question and option representations cross-attend to one another and back to the input payload, so that all questions in the schema are scored jointly rather than independently. A lexical prior derived from the option descriptions helps preserve the semantic meaning of each option. The head outputs one logit $z_{ji}$ for each option $a_{ji}\in A_j$, and the probabilities in Eq.~\ref{eq:decision_probability} are obtained through a per-question softmax:
\begin{equation}
P(a_{ji}\mid x,q_j)=\frac{\exp(z_{ji})}{\sum_{a_{jl}\in A_j}\exp(z_{jl})}.
\end{equation}

The model developers describe post-training with label-smoothed cross-entropy and a Brier loss, followed by reinforcement learning for calibrated decisions \cite{cloudflare2026clef}. This development process is distinct from the present evaluation, which performs inference without updating the released model parameters.

\definecolor{clefMain}{RGB}{214,96,46}    % CLEF accent (orange)
\definecolor{clefFlash}{RGB}{38,110,180}  % CLEF-Flash accent (blue)
\definecolor{inGray}{RGB}{245,245,245}
\definecolor{frozenGray}{RGB}{225,229,236}
\definecolor{headGreen}{RGB}{224,240,226}
\definecolor{headEdge}{RGB}{60,140,80}

\tikzset{
  clefbox/.style   = {draw, rounded corners=2pt, align=center, font=\footnotesize,
                      inner sep=3pt, line width=0.6pt},
  inbox/.style     = {clefbox, fill=inGray, minimum width=2.5cm},
  frozen/.style    = {clefbox, fill=frozenGray},
  head/.style      = {clefbox, fill=headGreen, draw=headEdge, text width=3.2cm},
  arr/.style       = {-{Stealth[length=5pt]}, line width=0.7pt},
  lbl/.style       = {font=\scriptsize\itshape, text=black!70},
}

% \ClefArch{<title>}{<backbone name>}{<params>}{<accent colour>}
\newcommand{\ClefArch}[4]{%
\begin{tikzpicture}[node distance=4mm and 7mm]
  % ---------------- Inputs ----------------
  \node[inbox, minimum height=1.25cm] (img) {Ultrasound image};
  \node[inbox, below=of img] (state) {State $x$\\{\scriptsize text / JSON (optional)}};
  \node[inbox, below=of state, text width=2.4cm] (schema) {%
    Schema $Q$ ($q_j$, \texttt{choice})\\[1pt]
    {\scriptsize $A_j{=}\{$\texttt{normal}, \texttt{benign},\\ \texttt{malignant}$\}$}};
  \node[lbl, above=1mm of img] (inlbl) {Input record};

  % ---------------- Encoders ----------------
  \node[frozen, right=of img, minimum width=2.2cm, minimum height=1.25cm] (vit)
       {Vision encoder\\{\scriptsize (from backbone)}};
  \node[frozen, right=of schema, minimum width=2.2cm] (tok)
       {Tokenizer\\{\scriptsize serialized record}};
  \coordinate (stmid) at ($(state.east)!0.5!(state.east -| tok.west)$);

  % ---------------- Token sequence ----------------
  \node[clefbox, fill=white, right=8mm of $(vit.east)!0.5!(tok.east)$,
        minimum height=3.6cm, minimum width=0.9cm, text width=0.75cm] (seq)
       {\rotatebox{90}{\scriptsize visual + text tokens}};

  % ---------------- Backbone ----------------
  \node[frozen, draw=#4, line width=1.1pt, right=of seq,
        minimum height=3.6cm, text width=2.6cm] (bb)
       {\textbf{#2}\\[2pt] {\scriptsize #3 params}\\[4pt]
        {\scriptsize fixed parameters}\\[4pt]
        {\scriptsize \textbf{prefill-only}}\\{\scriptsize single forward pass}};
  \node[lbl, above=1mm of bb] {Backbone (no decoding)};

  % ---------------- Hidden states ----------------
  \node[clefbox, fill=white, right=of bb, minimum height=1.0cm] (H)
       {$\mathbf{H}\in\mathbb{R}^{T\times d}$\\{\scriptsize final hidden states}};

  % ---------------- Joint schema head ----------------
  \node[head, right=9mm of H, yshift=8mm] (s1)
       {\textbf{Stage 1}\\{\scriptsize option $\rightarrow$ state}\\{\scriptsize evidence routing}};
  \node[head, below=4mm of s1] (s2)
       {\textbf{Stage 2}\\{\scriptsize cross-field attention}\\{\scriptsize \mbox{options $\leftrightarrow$ questions} \mbox{$\leftrightarrow$ input}}};
  \node[head, below=4mm of s2] (score)
       {Schema-bound scoring\\{\scriptsize + lexical prior}};
  \begin{scope}[on background layer]
    \node[draw=headEdge, dashed, rounded corners=4pt, fill=headGreen!35,
          inner sep=4pt, fit=(s1)(s2)(score)] (headbox) {};
  \end{scope}
  \node[lbl, above=0.5mm of headbox] {Joint schema head};

  % ---------------- Output ----------------
  \node[clefbox, fill=white, right=8mm of s2, text width=2.0cm] (soft)
       {logits $z_{ji}$\\[2pt] {\scriptsize per-question}\\ softmax};
  \node[right=7mm of soft, font=\scriptsize, anchor=west] (probs) {%
    \begin{tikzpicture}[x=1.4cm,y=0.32cm]
      \foreach \p/\n/\y in {0.08/normal/2, 0.22/benign/1, 0.70/malignant/0}{
        \fill[#4!80] (0,\y) rectangle (\p,\y+0.7);
        \draw[black!40] (0,\y) rectangle (1,\y+0.7);
        \node[anchor=east, font=\scriptsize] at (0,\y+0.35) {\texttt{\n}};
      }
      \node[font=\scriptsize] at (0.5,3.3) {$P(a_{ji}\mid x,q_j)$};
      \node[font=\scriptsize] at (0.5,-0.7) {$\hat a_j=\arg\max$};
    \end{tikzpicture}};

  % ---------------- Arrows ----------------
  \draw[arr] (img) -- (vit);
  \draw[arr] (schema) -- (tok);
  \draw[arr] (state.east) -| ($(tok.north west)!0.3!(tok.north east)$);
  \draw[arr] (vit.east) -- (vit.east -| seq.west);
  \draw[arr] (tok.east) -- (tok.east -| seq.west);
  \draw[arr] (seq) -- (bb);
  \draw[arr] (bb) -- (H);
  \draw[arr] (H.east) -- ++(3mm,0) |- (s1.west);
  \draw[arr] (H.east) -- ++(3mm,0) |- (s2.west);
  \draw[arr] (s1) -- (s2);
  \draw[arr] (s2) -- (score);
  \draw[arr] (score.east) -| ($(soft.south)$);
  \draw[arr] (soft) -- (probs);

  % ---------------- Title ----------------
  \node[font=\small\bfseries, text=#4, anchor=south west]
       at ($(inlbl.north west)+(0,1mm)$) {#1};
\end{tikzpicture}%
}

% ---------------- Figure ----------------
\begin{figure*}[t]
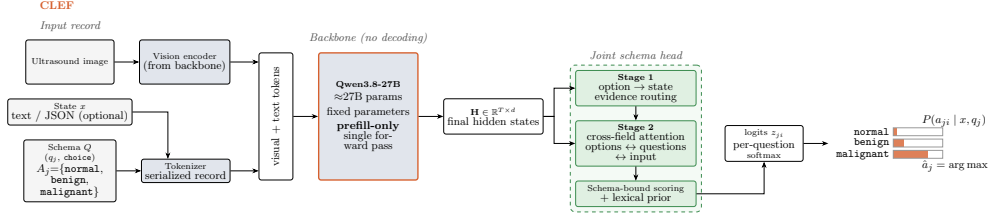
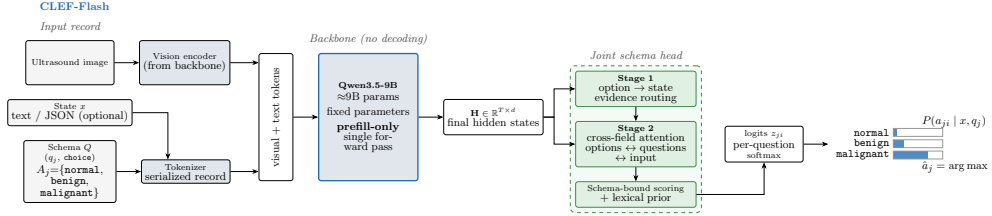

  \centering
  \begin{subfigure}{\textwidth}
    \centering
    \resizebox{\textwidth}{!}{\ClefArch{CLEF}{Qwen3.8-27B}{$\approx$27B}{clefMain}}
    \caption{CLEF ($\approx$27B parameters).}
    \label{fig:clef_arch}
  \end{subfigure}
  \\[4mm]
  \begin{subfigure}{\textwidth}
    \centering
    \resizebox{\textwidth}{!}{\ClefArch{CLEF-Flash}{Qwen3.5-9B}{$\approx$9B}{clefFlash}}
    \caption{CLEF-Flash ($\approx$9B parameters).}
    \label{fig:clefflash_arch}
  \end{subfigure}
  \caption{Architecture of the CLEF decision models applied to breast ultrasound classification.
  The image is encoded by the backbone's vision encoder, while the state and the typed decision schema
  are serialized as text. The multimodal backbone, held fixed during evaluation, performs a single
  prefill-only pass, and the joint schema head routes evidence from the input to each option,
  applies cross-field attention, and outputs one logit per admissible option. A per-question softmax
  yields the probability distribution over \texttt{normal}, \texttt{benign} and \texttt{malignant}.
  The panels show the respective backbones within the shared decision-model design. The three-class output illustrates BUSI; binary tasks use only benign and malignant options. Probabilities shown are illustrative.}
  \label{fig:clef_architectures}
\end{figure*}

\section{Materials and Methods}
\label{sec:methods}

\subsection{Experimental Design}
\label{subsec:design}

This study evaluates inference using the publicly released pretrained weights of CLEF and CLEF-Flash. Neither model is trained, fine tuned, or adapted on the evaluated datasets; no task specific classification head is introduced, and all model parameters remain unchanged.

\begin{figure}[h]
\centering

\begin{tikzpicture}[
    font=\footnotesize,
    >=Latex,
    dataset/.style={
        draw,
        rounded corners,
        align=center,
        minimum width=2.5cm,
        minimum height=0.55cm
    },
    imagebox/.style={
        draw,
        rounded corners,
        align=center,
        minimum width=3.2cm,
        minimum height=0.8cm
    },
    modelbox/.style={
        draw,
        rounded corners,
        align=center,
        minimum width=1.9cm,
        minimum height=0.8cm
    },
    promptbox/.style={
        draw,
        rounded corners,
        align=center,
        minimum width=0.8cm,
        minimum height=0.5cm
    },
    outputbox/.style={
        draw,
        rounded corners,
        align=center,
        minimum width=3.6cm,
        minimum height=0.8cm
    },
    evalbox/.style={
        draw,
        rounded corners,
        align=center,
        minimum width=4.0cm,
        minimum height=0.8cm
    },
    flow/.style={
        -{Latex[length=2mm]},
        line width=0.7pt
    },
    connector/.style={
        line width=0.7pt
    }
]

% ==========================
% Datasets
% ==========================

\node[dataset] (datasetb) {Dataset B};

\node[dataset,
      below=0mm of datasetb]
      (busi) {BUSI};

\node[dataset,
      below=0mm of busi]
      (busbra) {BUS-BRA};

% ==========================
% Original images
% ==========================

\node[imagebox,
      below=7mm of busbra]
      (images)
      {\textbf{Original Images}\\No Preprocessing};

\draw[flow]
      (busbra.south)
      --
      (images.north);

% ==========================
% Models
% ==========================

\node[modelbox,
      below left=11mm and 8mm of images]
      (flash)
      {\textbf{CLEF-Flash}\\9B};

\node[modelbox,
      below right=11mm and 8mm of images]
      (clef)
      {\textbf{CLEF}\\27B};

\coordinate (modelsplit)
      at ($(images.south)+(0,-4mm)$);

\draw[connector]
      (images.south)
      --
      (modelsplit);

\draw[flow]
      (modelsplit)
      -|
      (flash.north);

\draw[flow]
      (modelsplit)
      -|
      (clef.north);

% ==========================
% CLEF-Flash prompts
% ==========================

\node[promptbox,
      below=8mm of flash,
      xshift=-10mm]
      (fp1) {P1};

\node[promptbox,
      below=8mm of flash]
      (fp2) {P2};

\node[promptbox,
      below=8mm of flash,
      xshift=10mm]
      (fp3) {P3};

\coordinate (flashsplit)
      at ($(flash.south)+(0,-3mm)$);

\draw[connector]
      (flash.south)
      --
      (flashsplit);

\draw[flow]
      (flashsplit)
      -|
      (fp1.north);

\draw[flow]
      (flashsplit)
      --
      (fp2.north);

\draw[flow]
      (flashsplit)
      -|
      (fp3.north);

% ==========================
% CLEF prompts
% ==========================

\node[promptbox,
      below=8mm of clef,
      xshift=-10mm]
      (cp1) {P1};

\node[promptbox,
      below=8mm of clef]
      (cp2) {P2};

\node[promptbox,
      below=8mm of clef,
      xshift=10mm]
      (cp3) {P3};

\coordinate (clefsplit)
      at ($(clef.south)+(0,-3mm)$);

\draw[connector]
      (clef.south)
      --
      (clefsplit);

\draw[flow]
      (clefsplit)
      -|
      (cp1.north);

\draw[flow]
      (clefsplit)
      --
      (cp2.north);

\draw[flow]
      (clefsplit)
      -|
      (cp3.north);

% ==========================
% Prompt groups
% ==========================

\node[
      draw,
      dashed,
      rounded corners,
      inner sep=3pt,
      fit=(fp1)(fp2)(fp3)
      ] {};

\node[
      draw,
      dashed,
      rounded corners,
      inner sep=3pt,
      fit=(cp1)(cp2)(cp3)
      ] {};

% ==========================
% Merge predictions
% ==========================

\node[outputbox,
      below=17mm of $(fp2)!0.5!(cp2)$]
      (prediction)
      {\textbf{Diagnostic Prediction}\\
      Normal / Benign / Malignant};

\coordinate (merge)
      at ($(prediction.north)+(0,4mm)$);

\draw[connector]
      (fp1.south)
      --
      ++(0,-5mm)
      -|
      (merge);

\draw[connector]
      (fp2.south)
      --
      ++(0,-5mm)
      -|
      (merge);

\draw[connector]
      (fp3.south)
      --
      ++(0,-5mm)
      -|
      (merge);

\draw[connector]
      (cp1.south)
      --
      ++(0,-5mm)
      -|
      (merge);

\draw[connector]
      (cp2.south)
      --
      ++(0,-5mm)
      -|
      (merge);

\draw[connector]
      (cp3.south)
      --
      ++(0,-5mm)
      -|
      (merge);

\draw[flow]
      (merge)
      --
      (prediction.north);

% ==========================
% Evaluation
% ==========================

\node[evalbox,
      below=7mm of prediction]
      (evaluation)
      {\textbf{Classification Evaluation}\\
      Confusion Matrix and Classification Metrics};

\draw[flow]
      (prediction.south)
      --
      (evaluation.north);

\end{tikzpicture}

\caption{Overview of the inference and evaluation pipeline across datasets, models, and prompting strategies. The available diagnostic categories depend on the dataset, as specified in Table~\ref{tab:datasets}.}
\label{fig:experimental_flow}

\end{figure}
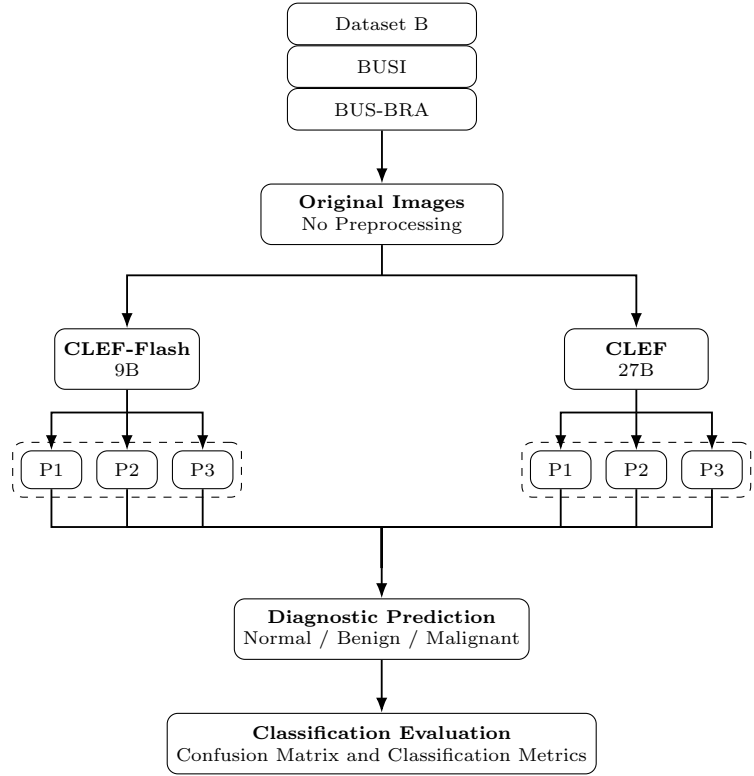

The experimental design crosses three datasets, two decision models, and three prompting strategies, yielding 18 dataset-model-prompt configurations. Figure~\ref{fig:experimental_flow} summarizes the evaluation pipeline.

The original dataset images are used without user-defined resizing, cropping, enhancement, denoising, normalization, segmentation, region-of-interest extraction, or data augmentation. Each image file is Base64-encoded for the inference request payload. This describes processing before submission; it does not imply that the inference engine's native image processor performs no internal transformations.

Ground truth labels are used only after inference for performance evaluation and are never included in the model input.

\subsection{Datasets}
\label{subsec:datasets}

We use Dataset B \cite{yap2018}, BUSI \cite{aldhabyani2020}, and BUS-BRA \cite{gomezflores2024}. Dataset B and BUS-BRA provide binary benign-malignant tasks, while BUSI additionally includes normal images. The evaluated sets contain 163, 780, and 1875 images, respectively, for a total of 2818 images. Table~\ref{tab:datasets} reports the class counts used in this evaluation. Metrics are computed separately for each dataset.

\begin{table}[h]
\centering
\caption{Summary of the breast ultrasound datasets used in this study.}
\label{tab:datasets}
\begin{tabular}{lcccc}
\hline
\hline
Dataset & Normal & Benign & Malignant & Task \\
\hline
\hline
Dataset B & -- & 109 & 54 & Binary \\
BUSI      & 133 & 437 & 210 & Three class \\
BUS-BRA   & -- & 1268 & 607 & Binary \\
\hline
\hline
\end{tabular}
\end{table}

\subsection{Inference and Prediction}
\label{subsec:models}

Inference was performed on a single NVIDIA A100 GPU with 80~GB of VRAM using Ollama version 0.35.1 as the inference engine, with FP16 precision configured. The observed GPU memory usage was 22,364~MiB for CLEF (27B) and 13,741~MiB for CLEF-Flash (9B). Each request contained the Base64-encoded original image and the corresponding prompt and decision schema.

For each image, a single classification question is defined using the \texttt{choice} decision type. The available choices correspond exactly to the diagnostic labels of the evaluated dataset. The model returns a probability for each allowed category, and the category with the highest probability is taken as the predicted class.

\begin{equation}
\hat{y} = \arg\max_{c \in C} P(c \mid I, P),
\end{equation}

where $I$ denotes the input ultrasound image, $P$ represents the prompting strategy, and $C$ denotes the set of diagnostic categories.

\subsection{Prompting Strategies}
\label{subsec:prompts}

The prompts vary the diagnostic instructions and category descriptions. P1 supplies the category names, P2 explains the diagnostic task, and P3 adds sonographic criteria (Table~\ref{tab:prompts}). Appendix~\ref{app:prompts} provides the three-class BUSI request examples. All examples already identify the input as a breast ultrasound image in the shared \texttt{state} field; P2 therefore elaborates the classification request rather than introducing the imaging modality. Binary tasks restrict the answer space to benign and malignant.

\begin{table}[h]
\centering
\caption{Prompting strategies evaluated in this study.}
\label{tab:prompts}
\begin{tabular}{p{0.12\linewidth}p{0.78\linewidth}}
\hline
\hline
Prompt & Description \\
\hline
\hline
P1 & Minimal classification instruction containing only the task and diagnostic alternatives. \\

P2 & Task-guided instruction explaining the diagnostic classification task. \\

P3 & Clinical knowledge guided instruction incorporating relevant sonographic characteristics such as lesion shape, margin, orientation, echogenicity, and posterior acoustic features. \\
\hline
\hline
\end{tabular}
\end{table}

\subsection{Evaluation Metrics}
\label{subsec:metrics}

We assess classification performance using accuracy, F1 score, balanced accuracy, and precision, with the breast ultrasound evaluation study of Tan et al.~\cite{TAN2026114134} as a related reference. Accuracy is the proportion of correctly classified images. For each class $c$, precision is $P_c=TP_c/(TP_c+FP_c)$, recall is $R_c=TP_c/(TP_c+FN_c)$, and F1 score is their harmonic mean, $F1_c=2P_cR_c/(P_c+R_c)$, where $TP_c$, $FP_c$, and $FN_c$ denote true positives, false positives, and false negatives under a one-versus-rest interpretation. Balanced accuracy is the mean recall across the $K$ classes, $K^{-1}\sum_{c=1}^{K}R_c$. Precision and F1 score are reported as unweighted means of their class-specific values (macro averages), and all four classification metrics are expressed as percentages.

Inference efficiency is summarized by frames per second (FPS), the number of images processed per second, and latency, the mean elapsed time per image in milliseconds. For $N$ images processed sequentially over a measured inference interval of $T$ seconds, $\mathrm{FPS}=N/T$ and mean latency is $1000T/N$ ms/image. Higher classification scores and FPS indicate better performance, whereas lower latency indicates faster inference.

\section{Results}
\label{sec:results}

Table~\ref{tab:results_all} summarizes classification performance and inference speed across the three datasets, three prompting strategies, and two models.
\begin{table}[htbp]
\centering
\caption{Classification performance and inference speed on Dataset B, BUSI, and BUS-BRA. Acc.: accuracy; Bal. Acc.: balanced accuracy. F1 score and precision are macro averages. FPS is reported in frames per second and latency in milliseconds per image. Bold indicates the best value for each metric within each dataset (higher is better except for latency).}
\label{tab:results_all}
\small
\setlength{\tabcolsep}{4pt}
\begin{tabular*}{\textwidth}{@{\extracolsep{\fill}}llcccccc@{}}
\hline
\hline
Prompt & Model & \shortstack{Acc.\\(\%)} & \shortstack{F1 score\\(\%)} & \shortstack{Bal. Acc.\\(\%)} & \shortstack{Precision\\(\%)} & FPS & \shortstack{Latency\\(ms/image)} \\
\hline
\hline
\multicolumn{8}{l}{\textbf{Dataset B}} \\
\hline
\multirow{2}{*}{P1} & CLEF & 70.55 & 68.76 & 70.51 & 68.54 & 0.125 & 7971.36 \\
 & CLEF-Flash & 73.62 & 71.74 & 73.27 & 71.29 & 0.414 & 2415.63 \\
\addlinespace[2pt]
\multirow{2}{*}{P2} & CLEF & 65.03 & 64.12 & 67.31 & 65.35 & 0.126 & 7962.04 \\
 & CLEF-Flash & 73.01 & 71.37 & 73.28 & 71.04 & 0.430 & 2324.27 \\
\addlinespace[2pt]
\multirow{2}{*}{P3} & CLEF & 65.03 & 64.95 & 71.52 & 70.23 & 0.126 & 7959.96 \\
 & CLEF-Flash & \textbf{74.85} & \textbf{73.70} & \textbf{76.52} & \textbf{73.65} & \textbf{0.458} & \textbf{2185.08} \\
\hline

\multicolumn{8}{l}{\textbf{BUSI}} \\
\hline
\multirow{2}{*}{P1} & CLEF & 65.77 & 63.11 & 65.79 & 66.19 & 0.161 & 6220.43 \\
 & CLEF-Flash & 62.56 & 59.45 & 62.77 & \textbf{68.98} & 0.297 & 3365.08 \\
\addlinespace[2pt]
\multirow{2}{*}{P2} & CLEF & 65.13 & \textbf{64.51} & \textbf{68.72} & 67.42 & 0.161 & 6221.19 \\
 & CLEF-Flash & 64.10 & 60.20 & 63.12 & 66.54 & \textbf{0.307} & \textbf{3258.00} \\
\addlinespace[2pt]
\multirow{2}{*}{P3} & CLEF & 59.87 & 58.83 & 64.23 & 65.59 & 0.161 & 6221.65 \\
 & CLEF-Flash & \textbf{66.28} & 63.49 & 66.60 & 67.85 & 0.306 & 3272.62 \\
\hline

\multicolumn{8}{l}{\textbf{BUS-BRA}} \\
\hline
\multirow{2}{*}{P1} & CLEF & 47.04 & 46.17 & 58.83 & 62.22 & 0.277 & 3609.67 \\
 & CLEF-Flash & 48.75 & 48.12 & 60.09 & 63.14 & 0.361 & 2766.75 \\
\addlinespace[2pt]
\multirow{2}{*}{P2} & CLEF & 45.12 & 43.56 & 58.22 & 63.58 & 0.264 & 3781.84 \\
 & CLEF-Flash & 50.08 & 49.66 & 60.86 & 63.32 & 0.361 & 2768.17 \\
\addlinespace[2pt]
\multirow{2}{*}{P3} & CLEF & 40.32 & 37.04 & 55.23 & 63.26 & 0.264 & 3782.60 \\
 & CLEF-Flash & \textbf{51.63} & \textbf{51.39} & \textbf{61.79} & \textbf{63.64} & \textbf{0.364} & \textbf{2746.68} \\
\hline
\hline
\end{tabular*}
\end{table}

\subsection{Classification Performance and Inference Speed}

On Dataset B, CLEF-Flash with P3 achieves the highest accuracy (74.85\%), macro F1 score (73.70\%), balanced accuracy (76.52\%), and macro precision (73.65\%). It also has the highest throughput (0.458 FPS) and lowest latency (2185.08 ms/image). CLEF-Flash exceeds CLEF in all four classification metrics under each matched prompt.

On BUSI, the best configuration depends on the metric. CLEF-Flash with P3 achieves the highest accuracy (66.28\%), whereas CLEF with P2 achieves the highest macro F1 score (64.51\%) and balanced accuracy (68.72\%). CLEF-Flash with P1 yields the highest macro precision (68.98\%), and P2 gives its highest throughput (0.307 FPS) and lowest latency (3258.00 ms/image). Thus, no configuration leads across all metrics on the three class task.

On BUS-BRA, CLEF-Flash with P3 leads all six metrics, but its accuracy and macro F1 score reach only 51.63\% and 51.39\%, respectively. Its balanced accuracy is 61.79\%, macro precision is 63.64\%, and latency is 2746.68 ms/image. CLEF-Flash has lower latency than CLEF under every matched prompt on all three datasets.

\subsection{Class-Specific Error Patterns}

Figures~\ref{fig:cm_busb}-\ref{fig:cm_busbra} in Appendix~\ref{app:confusion_matrices} show the confusion matrices. On Dataset B, changing CLEF from P1 to P3 reduces malignant cases predicted as benign from 16 to 5, but increases benign cases predicted as malignant from 32 to 52. For CLEF-Flash, the corresponding counts change from 15 to 10 and from 28 to 31. On BUSI, CLEF-Flash with P3 correctly identifies 65 normal and 260 benign images, compared with 51 and 235 under P1, while correctly identified malignant images decrease from 202 to 192. CLEF with P2 identifies the most normal images (83 of 133), helping explain its stronger balanced accuracy.

On BUS-BRA, errors are dominated by benign images predicted as malignant. CLEF with P3 correctly classifies 592 of 607 malignant images but only 164 of 1268 benign images. CLEF-Flash with P3 correctly classifies 550 malignant and 418 benign images. The resulting malignant recall is 97.53\% for CLEF and 90.61\% for CLEF-Flash, while benign recall is only 12.93\% and 32.97\%, respectively.

\section{Discussion}
\label{sec:discussion}

\subsection{Model and Prompt Dependence}

The results show that a larger model does not consistently provide better classification in this evaluation. CLEF-Flash outperforms CLEF under every matched prompt on the two binary datasets, while CLEF retains an advantage in macro F1 and balanced accuracy on BUSI with P2. Model selection therefore depends on both the dataset and the evaluation criterion. These comparisons concern two released models with different backbones and do not isolate parameter count as the cause of their performance differences.

The effect of clinical instructions is also model dependent. Moving from P1 to P3 improves CLEF-Flash accuracy on all three datasets, although Dataset B shows a small decrease at P2. For CLEF, the same change reduces accuracy on every dataset. The confusion matrices show that P3 shifts CLEF toward malignant predictions, increasing malignant recall at the expense of benign recall. These observations support evaluating prompts empirically rather than assuming that additional clinical detail will improve every aspect of classification. They do not establish whether the models use the described sonographic features as intended.

\subsection{Dataset Variation and Class Imbalance}

BUS-BRA exposes a substantial limitation of direct inference. Given its 1268 benign images among 1875 images, predicting every image as benign would yield 67.63\% accuracy, exceeding every evaluated configuration. Such a constant prediction would nevertheless have only 50\% balanced accuracy, below the observed 55.23-61.79\%. Together, these comparisons indicate some class discrimination but an unfavorable allocation of predictions across classes. The high malignant recall of CLEF with P3 must therefore be interpreted alongside its very low benign recall.

Performance differences across datasets suggest sensitivity to the evaluated data source, but these experiments do not isolate its cause. Acquisition conditions, image composition, and label characteristics are possible contributors that require separate analysis. BUSI also uses three classes rather than two, so differences in its aggregate scores cannot be attributed solely to dataset shift.

\subsection{Inference Efficiency}

CLEF-Flash offers lower observed latency across all matched comparisons, with approximately 3.30-3.64-fold speedups on Dataset B, 1.85-1.91-fold on BUSI, and 1.30-1.38-fold on BUS-BRA. These measurements support its efficiency advantage in the reported runs, but latency remains above two seconds per image even for the fastest configuration. The reported measurements apply to the A100/Ollama configuration described in Section~\ref{subsec:models}. These latency estimates are specific to the reported environment.

\subsection{Scope and Limitations}

The present comparisons are descriptive: the table provides point estimates without confidence intervals or paired significance tests. Small differences, such as the 0.51 percentage-point accuracy gap between CLEF-Flash/P3 and CLEF/P1 on BUSI, should not be interpreted as established superiority. The aggregate evaluation also does not establish probability calibration or the reliability of confidence-based selection. These require predictions linked to reference labels and an explicitly verified confidence definition. The observed results concern image-level classification; they do not establish patient-level diagnostic performance or clinical utility. The evaluation does not determine whether related images appeared in the models' original training data, so unchanged parameters during this study should not be equated with demonstrated absence of prior data exposure.

No supervised ultrasound model, biomedical foundation model, or radiologist was evaluated under the same protocol. Consequently, the results characterize the two decision models rather than establish superiority over existing diagnostic approaches. The image-level confusion matrices also do not account for within-patient dependence where multiple images are available. Future evaluation should include matched baselines, patient-aware uncertainty estimates, and calibration analysis.

\section{Conclusion}
\label{sec:conclusion}

This study evaluated inference with CLEF and CLEF-Flash across three breast ultrasound datasets and three prompting strategies. CLEF-Flash with P3 achieved the highest accuracy on each dataset and led all reported metrics on Dataset B and BUS-BRA, while CLEF with P2 achieved the strongest macro F1 and balanced accuracy on BUSI. CLEF-Flash was consistently faster in the reported environment. Nevertheless, poor benign recall on BUS-BRA and the contrasting response of the two models to clinical instructions show that structured outputs alone do not ensure reliable diagnostic classification. The benchmark supports evaluating dataset-specific errors, prompt sensitivity, and inference cost together. Matched baselines and patient-level validation are needed before broader conclusions about clinical use can be drawn.

\section*{Data Availability}

This study analyzed three existing datasets: BUSI~\cite{aldhabyani2020}, Dataset B~\cite{yap2018}, and BUS-BRA~\cite{gomezflores2024}. Access details are provided in the original publications. Dataset B requires approval of a licence agreement; access and reuse of all datasets are subject to the respective data providers' terms.

\section*{Code Availability}

The designated repository for the study code is \url{https://github.com/CaoMinhh/clef-breast-ultrasound}.

\clearpage
\begin{appendices}

\section{Prompt Specifications}
\label{app:prompts}

Three-class BUSI request examples are provided below. The shared state identifies the imaging modality. P1 names the categories, P2 elaborates their diagnostic meaning, and P3 provides sonographic criteria. Binary tasks use only benign and malignant alternatives.

% =========================================================
% Prompt 1
% =========================================================
\subsection{P1: Minimal Label Prompt}

\begin{lstlisting}[language=json]
{
  "state": "Breast ultrasound image",
  "images": ["<base64-encoded-image>"],
  "questions": {
    "diagnosis": {
      "type": "choice",
      "instructions": "Choose exactly one category: normal, benign, malignant.",
      "criteria": {
        "normal": "Normal",
        "benign": "Benign",
        "malignant": "Malignant"
      }
    }
  }
}
\end{lstlisting}

% =========================================================
% Prompt 2
% =========================================================
\subsection{P2: Task Guided Prompt}

\begin{lstlisting}[language=json]
{
  "state": "Breast ultrasound image",
  "images": ["<base64-encoded-image>"],
  "questions": {
    "diagnosis": {
      "type": "choice",
      "instructions": "Classify the breast ultrasound image according to whether it shows no lesion, a benign lesion, or a malignant lesion.",
      "criteria": {
        "normal": "No lesion",
        "benign": "Benign lesion",
        "malignant": "Malignant lesion"
      }
    }
  }
}
\end{lstlisting}

% =========================================================
% Prompt 3
% =========================================================
\subsection{P3: Clinical Knowledge Guided Prompt}

\begin{lstlisting}[language=json]
{
  "state": "Breast ultrasound image",
  "images": ["<base64-encoded-image>"],
  "questions": {
    "diagnosis": {
      "type": "choice",
      "instructions": "Evaluate the breast ultrasound image using lesion morphology, margin, orientation, echogenicity, posterior acoustic features, and surrounding tissue appearance. Choose exactly one category.",
      "criteria": {
        "normal": "No focal suspicious lesion is identified.",
        "benign": "Features predominantly compatible with a benign lesion, such as an oval/round shape, circumscribed margin, and parallel orientation.",
        "malignant": "Features predominantly suspicious for malignancy, such as irregular shape, non-circumscribed margin, non-parallel orientation, or suspicious posterior acoustic features."
      }
    }
  }
}
\end{lstlisting}

\clearpage
\section{Confusion Matrices}
\label{app:confusion_matrices}
\setcounter{figure}{0}

\begin{figure}[htbp]
\centering
\includegraphics[width=\textwidth]{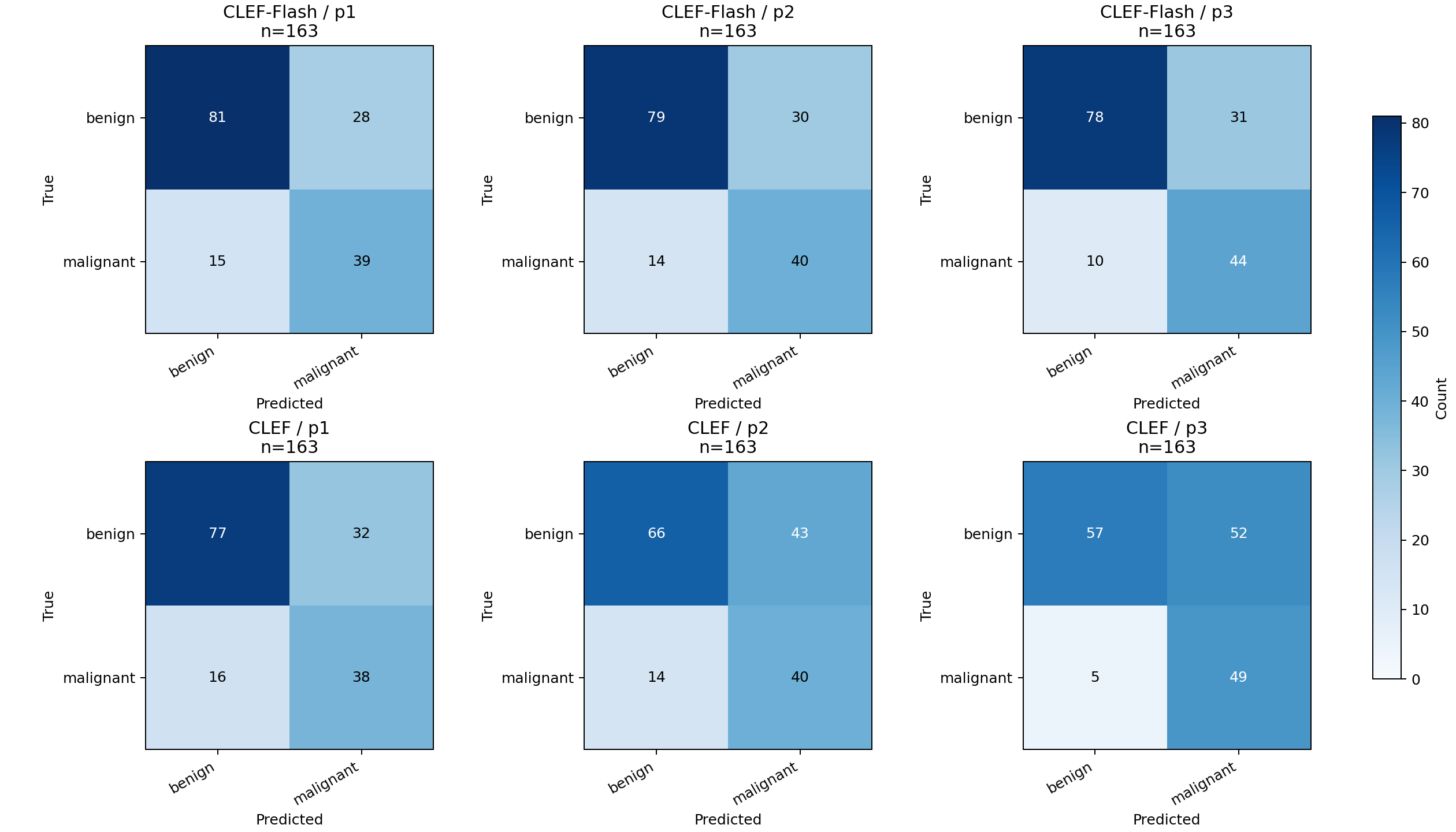}
\caption{Confusion matrices on Dataset B. Rows indicate reference labels and columns indicate predicted labels. The upper row shows CLEF-Flash and the lower row shows CLEF; columns correspond to P1, P2, and P3. Cell values are image counts.}
\label{fig:cm_busb}
\end{figure}

\begin{figure}[htbp]
\centering
\includegraphics[width=\textwidth]{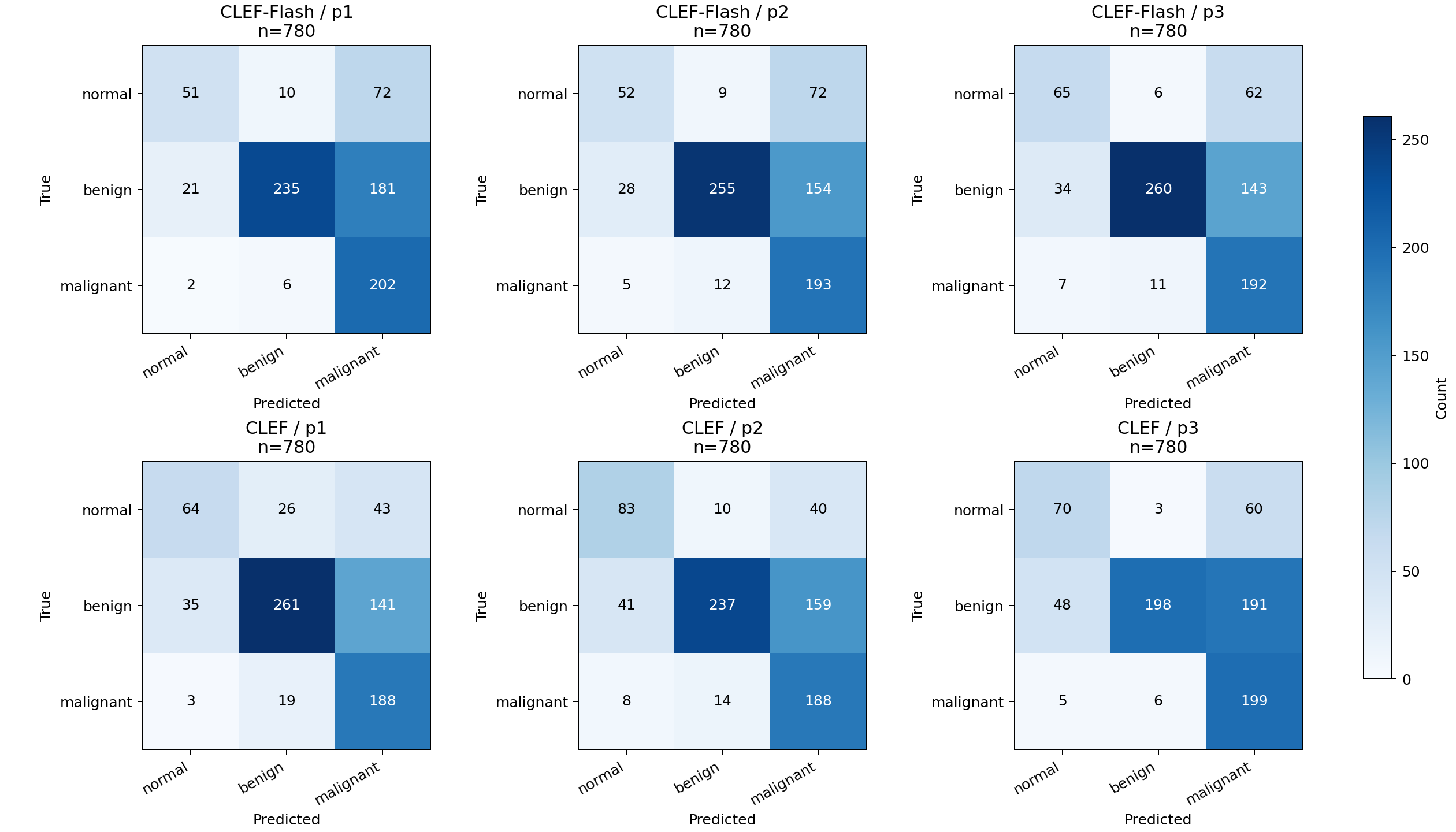}
\caption{Confusion matrices on BUSI. Rows indicate reference labels and columns indicate predicted labels. The upper row shows CLEF-Flash and the lower row shows CLEF; columns correspond to P1, P2, and P3. Cell values are image counts.}
\label{fig:cm_busi}
\end{figure}

\begin{figure}[htbp]
\centering
\includegraphics[width=\textwidth]{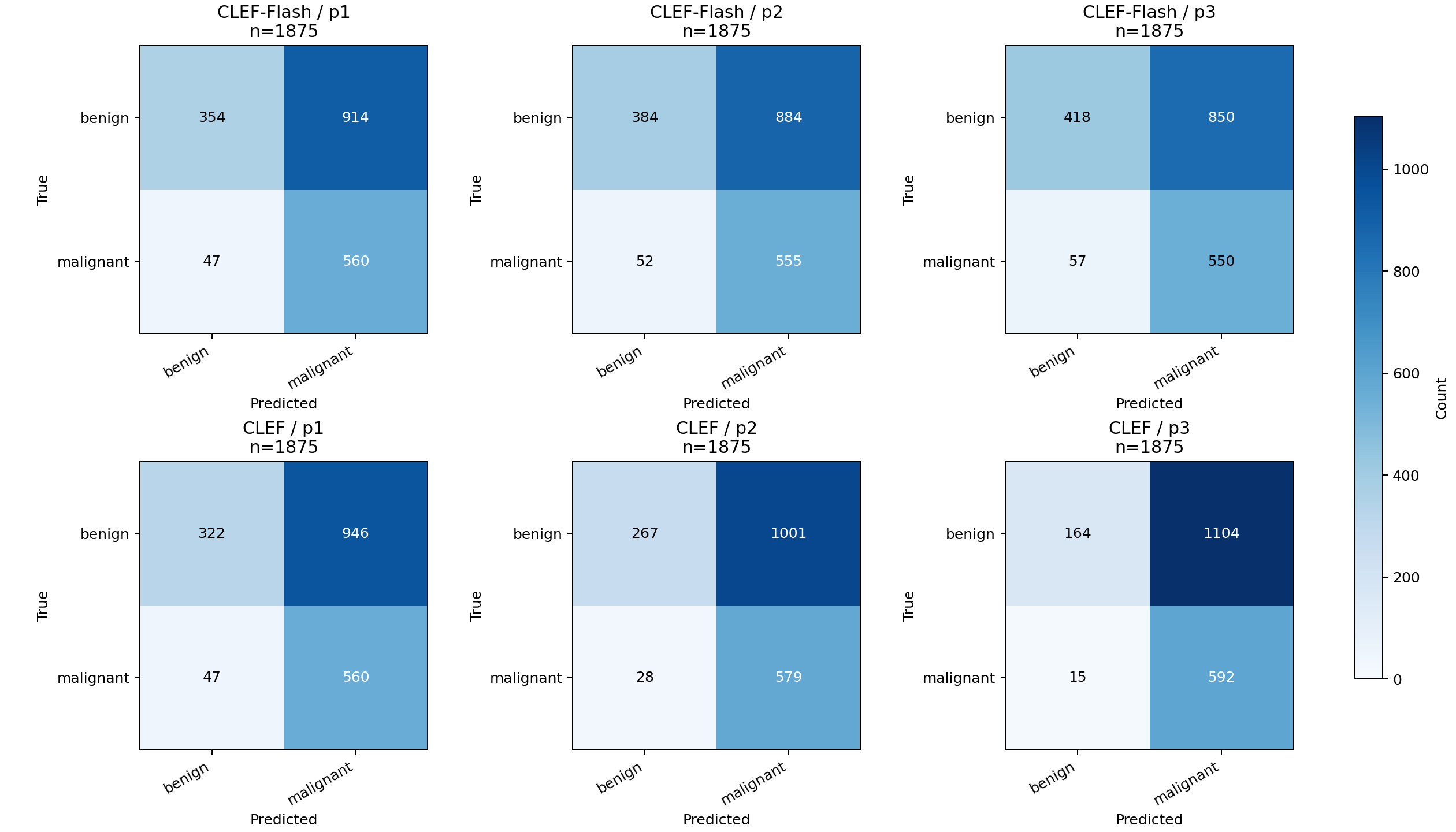}
\caption{Confusion matrices on BUS-BRA. Rows indicate reference labels and columns indicate predicted labels. The upper row shows CLEF-Flash and the lower row shows CLEF; columns correspond to P1, P2, and P3. Cell values are image counts.}
\label{fig:cm_busbra}
\end{figure}

\clearpage
\end{appendices}

\bibliography{reference}

\end{document}